\documentclass[aps,pra,twocolumn,superscriptaddress,nofootinbib]{revtex4-2}

\usepackage{dsfont}
\usepackage{xcolor,graphicx}
\usepackage{hyperref}
\usepackage{tabulary,graphicx}
\graphicspath{{Images/}}
\usepackage{diagbox}
\usepackage{amsmath,wasysym}
\usepackage{mathtools}
\usepackage{media9}
\usepackage{hyperref}

\newcommand{\ket}[1]{|{#1}\rangle}
\newcommand{\bra}[1]{\langle{#1}|}

\newcommand{\tr}{\text{Tr}}

\definecolor{LB}{RGB}{66,10,146}    

\begin{document}

\title{Quantum Noise Sensing by generating Fake Noise}

\author{Paolo Braccia} 
\address{Dipartimento di Fisica e Astronomia, Universit\`a di Firenze, I-50019, Sesto Fiorentino (FI), Italy}
\address{INFN, Sezione di Firenze, I-50019, Sesto Fiorentino (FI), Italy}
\address{LENS and QSTAR, Via N. Carrara 1, I-50019 Sesto Fiorentino, Italy.}

\author{Leonardo Banchi}
\address{Dipartimento di Fisica e Astronomia, Universit\`a di Firenze, I-50019, Sesto Fiorentino (FI), Italy}
\address{INFN, Sezione di Firenze, I-50019, Sesto Fiorentino (FI), Italy}

\author{Filippo Caruso}
\address{Dipartimento di Fisica e Astronomia, Universit\`a di Firenze, I-50019, Sesto Fiorentino (FI), Italy}
\address{LENS and QSTAR, Via N. Carrara 1, I-50019 Sesto Fiorentino, Italy.}
\address{Istituto Nazionale di Ottica CNR-INO, Firenze, Italy.}

\vspace{10pt}

\begin{abstract}
	Noisy-Intermediate-Scale-Quantum (NISQ) devices are nowadays starting to become available to the final user, hence potentially allowing to show the quantum speedups predicted by the quantum information theory. However, before implementing any quantum algorithm, it is crucial to have at least a partial or possibly full knowledge on the type and amount of noise affecting the quantum machine. Here, by generalizing quantum generative adversarial learning from quantum states (Q-GANs) to quantum operations/superoperators/channels (here named as {\sc SuperQGAN}s), we propose a very promising framework to characterize noise in a realistic quantum device, even in the case of spatially and temporally correlated noise (memory channels) affecting quantum circuits. The key idea is to learn about the noise by mimicking it in a way that one cannot distinguish between the real (to be sensed) and the fake (generated) one. We find that, when applied to the benchmarking case of Pauli channels, the {\sc SuperQGAN} protocol is able to learn the associated error rates even in the case of spatially and temporally correlated noise. Moreover, we also show how to employ it for quantum metrology applications. We believe our {\sc SuperQGAN}s pave the way for new hybrid quantum-classical machine learning protocols for a better characterization and control of the current and future unavoidably noisy quantum devices.
\end{abstract}

\pacs{03.67.Ac,03.67.Lx}

\maketitle

\section{Introduction}

The quest for a fully-operational, fault-tolerant quantum computer is still in its infancy. Running powerful, and possibly world-changing, quantum algorithms such as Shor's one \cite{ShorAlg} will still take some time, as a huge number of operative qubits is needed to implement error-correcting codes \cite{nielsen2010quantum}, which are very much needed because of the sensitivity to noise for almost all quantum protocols.\\
However, these years are nonetheless exciting for quantum computing, as they belong to the NISQ (Noisy Intermediate Scale Quantum) era \cite{preskill2018quantum}. Indeed, quantum processors of up to fifty qubits are actually available, and even though they are \textit{noisy} and \textit{small}, we can use them to look for proofs of principle of the coveted \textit{quantum supremacy} \cite{arute2019quantum}, driven by the observation that classical devices already are not able to simulate these processors. 
Besides the impossibility of running error-correcting protocols on such devices, due to their limited size, their effectiveness in delivering reliable quantum algorithms is doomed by the unavoidable interaction of the quantum system, realizing the computational register of qubits, with the external environment. This will probably be the biggest experimental challenge we will have to overcome in order to move on to the quantum era of computation. These unwanted couplings, on top of limiting the depth of the quantum circuits that can be reliably devised, may also induce back-flows of information from the environment to the computing system, leading to the observation of \textit{memory effects} when repeatedly using a given quantum gate. Characterizing the noise occurring on NISQ processors is then of great importance, as it can lead to devise tailored circuital schemes that can minimize error rates, or even exploit noisy processes to achieve the desired goal -- see for instance Refs. \cite{levy2017action,wang2021,caruso2010}.

\begin{figure}[t!]
    \centering
    \includegraphics[width=1.\linewidth]{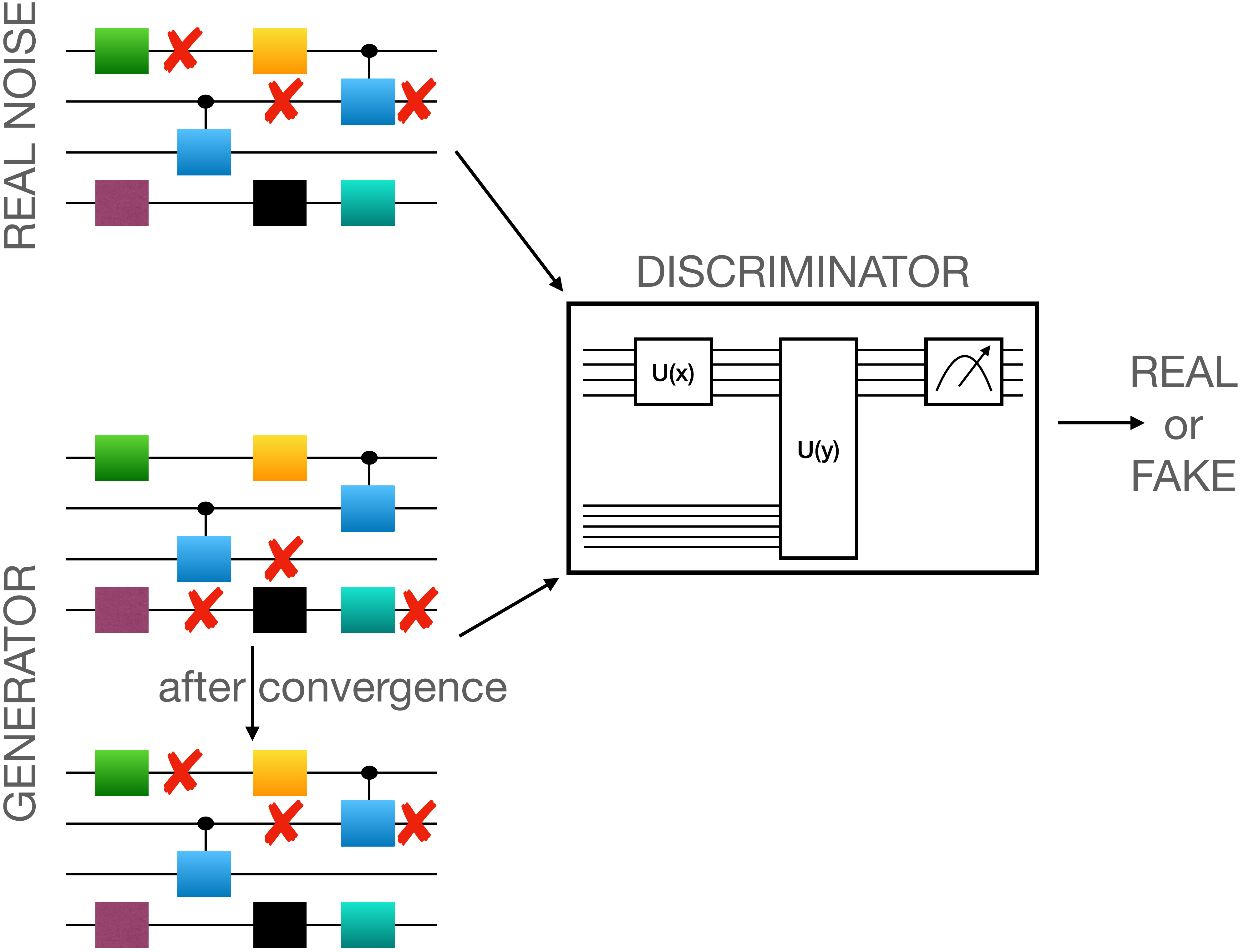}
    \caption{Pictorial representation of a {\sc SuperQGAN}, where the Discriminator needs to distinguish a real noisy quantum circuit from a fake one created by the Generator. These two agents play against each other, in particular the Generator needs to generate better and better data such that the task of the Discriminator becomes more and more complicated. The game ends (convergence) when the generator learns to create the real noisy quantum circuit (i.e., fake=real), hence identifying the errors occurring in the real circuit (crosses) running on a NISQ device. }
    \label{fig:superQGAN}
\end{figure}

In recent years, machine learning (ML) has overtaken the computational world, providing many powerful tools to tackle very complex tasks as domotic systems, autonomous cars, face/voice recognition, and medical diagnostics. It did not take long to realize that ML can be beneficial also to quantum computation, and many quantum adaptation of famous ML algorithms have been studied and discussed \cite{farhi2014quantum, pepper2019experimental, peruzzo2014variational,caruso2021}. As a matter of fact, a whole new branch of quantum computation, dubbed quantum machine learning (QML) \cite{biamonte2017quantum, lamata2020quantum}, has risen, exploiting the good behaviour of hybrid quantum-classical computational schemes to look for possible quantum advantages in ML tasks \cite{riste2017demonstration} and also to solve genuinely quantum problems \cite{lamata2020quantum}.
Moreover, machine learning methods have also been employed to learn quantum noise 
\cite{harper2020efficient,banchi2018modelling}.
Among the plethora of QML algorithms, quantum generative adversarial networks (QGANs) have shown great promise in generative tasks \cite{lloyd2018quantum, dallaire2018quantum,braccia2020enhance}, thanks to their ability to learn the properties of the quantum states they are faced with. 

In this work we show how to generalize the QGAN architecture from the context of quantum states to the context of quantum maps (or superoperators). In other words, the real data is represented by a real noisy quantum map while the generator creates quantum maps mimicking the real (unknown) one. We call them as {\sc SuperQGAN}s.

The paper is outlined as follows. In Sec. \ref{sec:sqgan_def} we introduce the mathematical definition of {\sc SuperQGAN}s and discuss their general setup. Then, in Sec. \ref{sec:examples} we first review the theory of Random Unitary Maps (\ref{sec:rum}) and later test our method against Pauli channels with spatial (\ref{sec:spatial}) and temporal (\ref{sec:temporal}) noise correlations. The Section ends with an application of the {\sc SuperQGAN} to a quantum metrology problem (\ref{sec:noise_metr}). Conclusions and outlooks are drawn in Sec. \ref{sec:conclusions}.

\section{Definition of {\sc SuperQGANs} for quantum maps}\label{sec:sqgan_def}

When dealing with experimental quantum processors, the circuital paradigm of perfect quantum computation remains an ideal abstraction. Indeed, the simple operations one would like to compose in order to build the desired algorithm are not perfect unitary evolutions of the targeted systems. Rather, they also induce unwanted, but also unavoidable, couplings with the environment leading, for example, to decoherence and loss of quantumness. It is thus more appropriate to address the physical processes occurring in a NISQ processor with the most general formalism of \textit{quantum operations} or \textit{quantum maps} \cite{nielsen2010quantum,caruso2014quantum}. This means that rather than associating a quantum circuit, or any of the gates constituting it, with a unitary $U$ mapping the input state as $\ket{\psi}\rightarrow U\ket{\psi}$, we have to represent it as a general CPTP map $\Phi$. The latter is a completely positive (CP) and trace-preserving (TP) linear super-operator acting on the space of density operators of the input system $\Phi : \rho\rightarrow\Phi(\rho)$. Notice that, when the input and output spaces are the same, they are also called as quantum channels.
When a single qubit map is independently applied (in parallel) to $n$ qubits, then the global
quantum map reads as $\Phi^{\otimes n}$. When this map is applied $n$ times (in series) to 
the same qubit, we will write it as $\Phi^n = \Phi\circ\cdots\circ\Phi$. In both cases, it is assumed that the noisy operations 
are uncorrelated: there are no spatial or temporal noise correlations. 

\begin{figure}[t!]
    \centering
		\includegraphics[width=1.\linewidth]{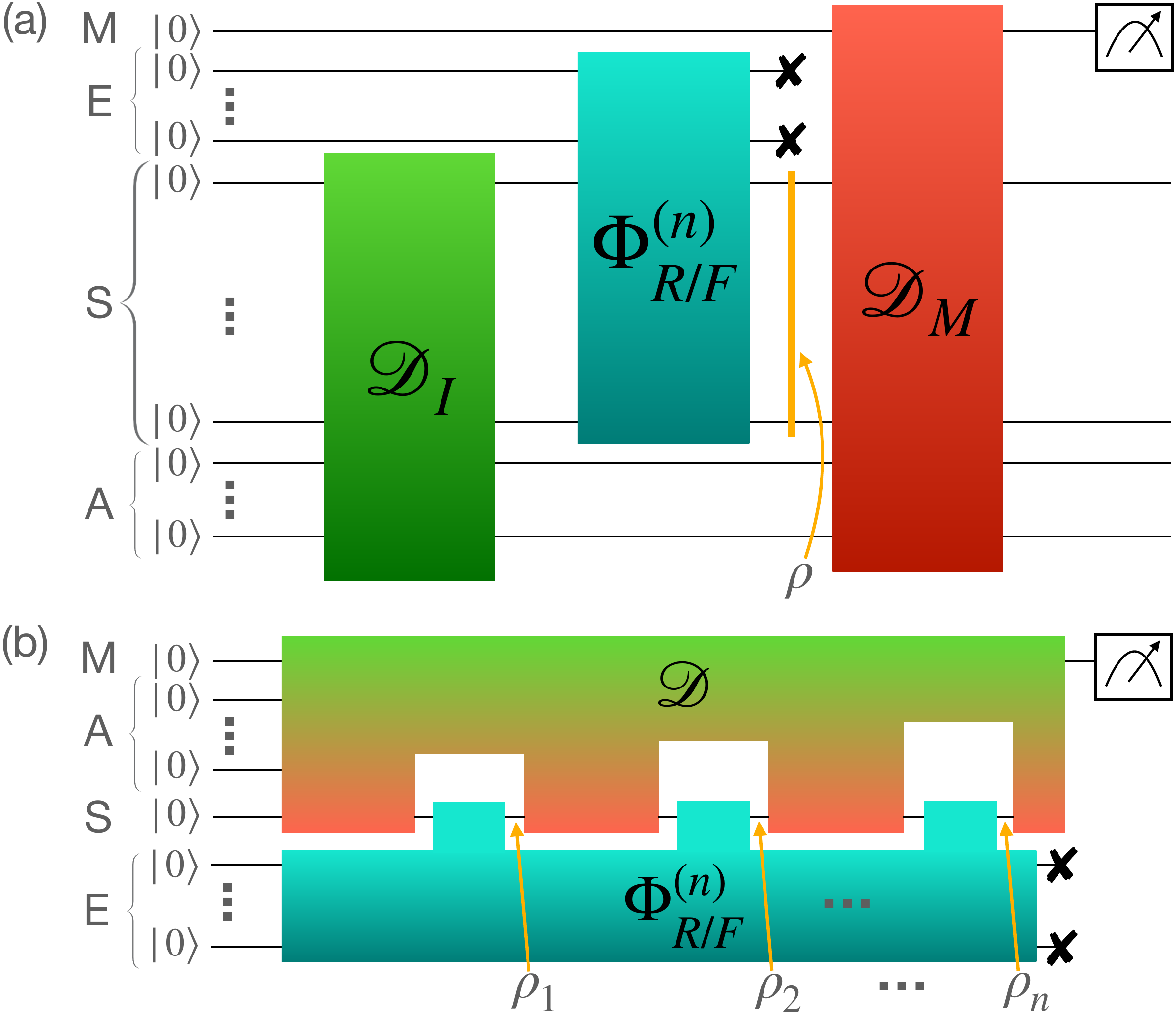}
		\caption{
			General detection scheme for spatially correlated (a) and temporally correlated (b) noise. 
			Noise couples the system qubits S with the environmental qubits E. We use the same diagram 
			to display both the real noise $\Phi_R^{(n)}$ and the generated fake noise $\Phi_F^{(n)}$, though 
			these may physically correspond to different evolutions -- e.g.~real interaction with an 
			environment vs.~a quantum circuit. 
			The discriminator has access to auxiliary qubits A and a measurement qubit M. Based on the 
			measurement outcome on M, the map $\Phi^{(n)}$ is judged either real or fake. 
			For spatially correlated noise (a), the generator applies an initialization map $\mathcal D_I$ 
			on S+A and a measurement map $\mathcal D_M$ on S+A+M, finally measuring M. 
			For temporally correlated noise (b), the discriminator applies the general map $\mathcal D$ that probes 
			the system S at intermediate times, finally measuring M. 
			In both cases, the discriminator has no access to the environmental qubits E. 
		}
    \label{fig:sens_qgan}
\end{figure}

However, in a NISQ device neighboring qubits typically experience spatially correlated noise, 
and the later-time evolution may display (non-Markovian) memory effects, hence leading to temporal noise correlations. As depicted in
Fig.~\ref{fig:sens_qgan}, both these cases can be represented by the action of
the map $\Phi^{(n)}$ that is much more general than either  $\Phi^{\otimes n}$ or $\Phi^n$. 
For spatially correlated noise, $\Phi^{(n)}$ maps $n$-qubit states to $n$-qubit states, while for
temporally correlated  noise $\Phi^{(n)}$ maps a single qubit to a ``history'' of single qubit states $\rho_t$, 
with $t=1,\dots,n$, each representing the state of the system at time $t$ -- see Fig.~\ref{fig:sens_qgan}.

A recent development of QML is the formalization of Quantum Generative Adversarial Networks (QGANs) \cite{lloyd2018quantum, dallaire2018quantum}, i.e. a generative model for quantum data. Mimicking the classical GAN scheme \cite{goodfellow2014generative}, QGANs work by exploiting an adversarial game where a Generator (G) agent, able to produce tunable fake instances of some target (real) distribution of data, is opposed to a Discriminator (D) that is in turn able to find good strategies to tell real and fake data apart. Played in turns, this game can be framed in Nash's game-theory and, under reasonable assumption of convexity, possesses a unique equilibrium point \cite{kakutani1941generalization}, where G is able to completely fool D and achieve a perfect data copying strategy. QGANs use quantum states to encode target and fake data, when the latter are classical \cite{romero2019variational, zoufal2019quantum, anand2020experimental}, or to directly assess the problem of reproducing the output of some unknown quantum process \cite{benedetti2019adversarial, braccia2020enhance}. Dubbing $\sigma$ the target quantum state and $\rho$ the generated one, the QGAN game can be expressed as the min-max game
$  \min_{\rho}\max_{M}\, S(\rho, M)$ where 
$S(\rho, M) = p(R|\sigma) - p(R|\rho)$
is called the \textit{score-function}, $p(R|\rho) =\tr[M\rho]$ is the
probability of judging real (R) the state $\rho$, and the operator
$M$ is part of a two-outcomes positive-operator-valued-measurement (POVM) $\Pi = \{M, I-M\}$. 
The QGAN score function is readily interpreted as the difference between the probability that D, controlling $M$, labels correctly $\sigma$ as real, and that of wrongly labelling $\rho$ as such. 

Without imposing any restriction on the set of possible measurements and
generated states, we can interpret the convergence of a QGAN as the completion
of an adaptive  state tomography protocol, and this motivates us to extend this
framework to tackle the more general \textit{quantum process tomography}.
We define a {\sc SuperQGAN} as the two-player game between a generator (G) and a discriminator (D),
where G tries to reproduce a general CPTP map $\Phi^{(n)}$ and D tries to distinguish the real map 
from the fake one. 
We focus on the general maps $\Phi^{(n)}$ introduced in Fig.~\ref{fig:sens_qgan} 
that describe, for instance, either spatially correlated or temporally correlated 
noise. For temporal correlations, the map can be expressed as a ``quantum comb'' 
\cite{chiribella2008quantum}. Quantum combs are graphical representations of quantum circuits 
that possibly couple the system to environmental ancillary qubits, as in Fig.~\ref{fig:sens_qgan}(b). 
Each ``comb tooth'' models quantum operations on the system at a certain time. The system state 
$\rho_{t}^{\rm in}$ goes into the tooth from the left and the output state $\rho_{t}^{\rm out}$ exits 
from the right. All teeth are linked via the comb shaft, which represents environmental memory effects due
to entanglement or other correlations between system and environment. 
Without external perturbations $\rho_t^{\rm in} = \rho_{t-1}^{\rm out}$, while in general the 
input and output states will be different if the system is probed, as in Fig.~\ref{fig:sens_qgan}(b).
For both spatial and temporal correlations,
the discriminator can use all the resources offered by quantum mechanics to 
optimally discriminate the two processes \cite{watrous2018theory}:  
these 
include entangling the system with a suitable number of ancillary qubits and performing
generalized measurements (POVM) on the extended space. Nonetheless, the discriminator has no access 
to the environment responsible for the noisy evolution (see Fig.~\ref{fig:sens_qgan}). 

The {\sc SuperQGAN} can be mathematically described as the following min-max game 
\begin{equation}
	\min_{\Phi_F^{(n)}}\max_{\mathcal D} \tr\left[\mathcal D\star\left(\Phi_R^{(n)}-\Phi_F^{(n)}\right)
	\left(\ket0\!\bra0^{\otimes N_D}\right)\right],
	\label{e:superqgan}
\end{equation}
where $N_D$ is the total number of qubits used by the discriminator, namely the sum 
of system qubits S, ancillary qubits A and measurement qubits M in Fig.~\ref{fig:sens_qgan}, 
$\Phi_{R/F}^{(n)}$ are, respectively, the real (R) and fake (F) process maps, while
$\mathcal D$ describes the set of operations performed by the discriminator. 
When the task is to discriminate between two processes as in Fig.~\ref{fig:sens_qgan}(a), then
$\mathcal D=(\mathcal D_I,\mathcal D_M)$ is a pair of CPTP maps, the initialization map $\mathcal D_I$ and the measurement 
map $\mathcal D_M$, and the {\it star-operation} in Eq.~\eqref{e:superqgan} refers to the composition map
$\mathcal D\star\mathcal E = \mathcal D_M\circ\mathcal E\circ\mathcal D_I$, as in 
Fig.~\ref{fig:sens_qgan}(a), with $\mathcal E$ being a CPTP map.
When the task is to discriminate between two quantum ``combs'', 
as in Fig.~\ref{fig:sens_qgan}(b), the discriminator's strategy can be entirely different:
the discriminator can probe the system at all times $t=1,\dots,n$ and, by doing this, alter the state in S. 
In other terms, the input $\rho_t^{\rm in}$ in the $t$-th comb tooth will be different from the output 
$\rho_{t-1}^{\rm out}$ from the previous tooth.
As a consequence, all the outputs in S at later times will be altered. The probe can be effectively implemented 
via measurements or via operations that couple the system S with the ancillary qubits A, owned by the discriminator.
All these operations are grouped into a process map $\mathcal D$, which is pictorially written via 
the ``upside-down comb'' in  Fig.~\ref{fig:sens_qgan}(b), while the combined action of $\mathcal D $ and $\Phi^{(n)}$ is 
represented by the {\it star-operation} in Eq.~\eqref{e:superqgan}. 

Another way to harness the adversarial game strategy with CPTP maps is to use a
QGAN to learn their associated Choi-Jamio{\l}kowski (CJ) states
\cite{watrous2018theory}. Indeed, there is an isomorphism between  CPTP maps
$\Phi$ acting over $D$-dimensional quantum systems and the bipartite states
$ C_\Phi = (\mathcal{I}\otimes \Phi)(\ket\Omega\!\bra\Omega)$ 
living in $D^2$-dimensional quantum systems,
where $\ket\Omega =  \sum_{i=1}^D \ket{i,i}/\sqrt  D$. 
Since $C_\Phi$ is generally a mixed state, a {\sc SuperQGAN} game can be mapped to
a QGAN for mixed (bipartite) states \cite{braccia2020enhance}. Formally this approach
corresponds to a particular discrimination strategy 
in the general {\sc SuperQGAN} game of Eq.~\eqref{e:superqgan}: 
we need as many ancillary qubits as the system qubits in
Fig.~\ref{fig:sens_qgan}(a), and we need to fix the initialization circuit $\mathcal D_I$ such that 
the input for CPTP map $\Phi^{(n)}$ is $\ket\Omega$, namely 
$\mathcal D_I(\ket{0}\!\bra0^{\otimes N_D}) = \ket\Omega\!\bra\Omega$. 

In the following applications, we will approximate the general maps $\Phi_F^{(n)}$ and $\mathcal D$ via quantum 
circuits with a certain depth and with a certain amount of ancillary qubits. For spatially 
correlated noise, without any restriction on the possible operations, 
the maximization over $\mathcal D$ in Eq.~\eqref{e:superqgan} results in the 
diamond distance \cite{watrous2018theory} between two channels 
$\big\|\Phi_R^{(n)}-\Phi_F^{(n)}\big\|_\diamond$, whose minimum is always zero with
$\Phi_R^{(n)} = \Phi_F^{(n)}$. On the other hand, when either D or G have access to 
non-universal resources, the final value in \eqref{e:superqgan} may be greater than 
zero and in general $\Phi_R^{(n)} \neq \Phi_F^{(n)}$.  For instance, a restricted discrimination 
strategy without ancillary qubits will be computationally simpler, yet not general 
enough. On the other hand, deep quantum circuits with many ancillary qubits may be universal, 
yet numerically hard to {\it train}.

\section{Examples}\label{sec:examples}

\subsection{Random Unitary Operations}\label{sec:rum}

A random unitary operation describes a physical process that can be decomposed into the probabilistic application of one of a finite set of unitary operations \cite{caruso2010}. It has been demonstrated that in this case if one has access to the environment introducing noise and can measure it obtaining classical information, then the corresponding noise process can be corrected \cite{werner2003}. In a real quantum computer this might be also the very likely case when one is dealing with a quantum circuit that is ideally a unitary transformation on some initial qubit states but in practice each gate of the circuit with some probability can correspond to a slightly different gate. Since the user has no access to such information, this introduce noise in the quantum computation that can be described by a random unitary map. Random operations can be also exploited to create quantum information scrambling as it was experimentally demonstrated in a 10-qubit trapped-ion quantum simulator \cite{zoller2020}. Moreover, random operations can also allow to tailor the noise for scalable quantum computation via randomized compiling 
\cite{emerson2016}. In Ref. \cite{sholten2019} they exploit ML to classify single-qubit stochastic errors that can be written as a convex combination of unitary operations \cite{sholten2019}.
From the mathematical point of view, these transformations are described by CPTP maps whose action on an input state $\rho$ can be put in the form 
\begin{equation}\label{random_unitary_maps}
	\Phi_R(\rho)= \int  U(s) \rho U(s)^\dagger p(s) ds
\end{equation}
where $p(s)$ is a probability density and $ U(s) $ are some unitary
operators. That is, a random unitary map implements a particular non-unitary evolution
of the system, where different unitary evolutions happen in a probabilistic manner. 
We focus on the simple case where the operators $U(s)$ are known and the task is to reconstruct the 
probability density $p(s)$. This task can be naturally expressed as {\sc SuperQGAN} 
where the cost function \eqref{e:superqgan} depend linearly on $p(s)$.
The generator G can use a trial CPTP map 
	$\Phi_F(\rho)= \int  U(s) \rho U(s)^\dagger q(s) ds$ 
where the unitary operators $U(s)$ are those entering in \eqref{random_unitary_maps}, assumed to be known,
while $q(s)$ must be learnt during the game. Even if the discriminator D can apply all possible detection schemes,
this game {\it can} end with $p(s)\neq q(s)$. Mathematically speaking, the mapping $p(s)\mapsto \Phi_R$ is not 
injective and we may get $\Phi_F = \Phi_R$ even with $p(s)\neq q(s)$. 
This possibility can be formally checked by 
studing the CJ state of the random unitary map $\Phi_R$, which is given by the convex combination $C_{\Phi_R} = \int C_{U(s)} p(s) ds$ of the CJ states $C_{U(s)}$ of the unitary channels $\rho\mapsto U(s)\rho U(s)^\dagger$, with the same probability density $p(s)$. 
In general the states $C_{U(s)}$ are not linearly independent and the perfect reconstruction of the random unitary 
map is not enough to learn $p(s)$. 

In what follows, we will study some relevant examples to see how different properties of the channel affect the 
complexity of reconstructing $p(s)$.

\subsection{Pauli channels: spatial correlations}\label{sec:spatial}

It is reasonable to assume that the average noise affecting a quantum circuit is a Pauli channel \cite{Knill2005}, which represents a very large family of random unitary maps. Although this class is not the more general one, one can show that a Pauli map can exceptionally well approximate any realistic noise without introducing new errors \cite{Wallman2016,Ware2011}.
Learning schemes for Pauli channels have been previously discussed in Refs. \cite{Flammia2020,Harper2020}. Their methods rely on acquiring a large dataset of $n$-qubits measurement results that later get analysed to efficiently infer the Pauli error rates \cite{Harper2020}, or an averaged version of those \cite{Flammia2020}. In contrast, our procedure will need only single qubit measurements and uses machine learning techniques to produce error rates that get closer to the real ones after each measurement.

More specifically, Pauli channels belong to the family of random unitary channels described before
in Eq.~\eqref{random_unitary_maps}, where the unitary operators $U(s)$ belong to the {\it discrete} set of Pauli matrices. 
In the single qubit case,
these are obtained by choosing $\{U(k)\}=\{\sigma_k\}$ with $\sigma_0 =
\mathrm{I}$ and $\sigma_{1:3}=\{X,Y,Z\}$, i.e. they are convex combinations of
Pauli evolutions. 
This single-qubit Pauli channel can be readily extended to the $n$-uses case, both in series (e.g., when the channel is applied $n$ times to the same qubit) and in parallel (e.g., when multiple copies of the channel are used to process a string of input qubit states at the same time). Let us now consider these spatial and temporal correlations separately. 

For spatial correlations, the channel can be represented as in 
Fig.~\ref{fig:sens_qgan}(a), and  maps $n$-qubit states $\rho^{(n)}$ to $n$-qubit states as follows
\begin{equation}\label{pauli_ch_n}
	\Phi_p^{(n)}(\rho^{(n)}) = \sum_{\Vec{k}} p_{\Vec{k}}^{(n)} \sigma^{(n)}_{\Vec{k}} \rho^{(n)} \sigma_{\Vec{k}}^{(n)},
\end{equation}
where $\sigma_{\Vec{k}}^{(n)}=\sigma_{k_1}\otimes\dotsc\otimes\sigma_{k_n}$ are  Pauli strings, and $\vec k$ is a multi-index. 
It is simple to check that the CJ states of different channels $C_{\sigma_{\vec k}}$ are linearly independent, so 
the mapping $p_{\Vec{k}}^{(n)} \mapsto 	\Phi_p^{(n)}$ is bijective. 
Notice that, when the probability can be factorized as $p_{\vec k}^{(n)}=\prod_{j=1}^{n}p(k_j)$, the channel 
has no spatial correlations, and $\Phi_p^{(n)}$ is a tensor product of independent channels on each qubit,
whereas the above factorization property does not hold anymore
when the noise is correlated. 
We can exploit the above relation to check for spatial correlations,
by first learning $p({k})$ and then using it to
check if $p_{\Vec{k}}^{(2)}$ is factorized or not. 
Indeed, it suffices to show $\Phi^{(2)}\neq \Phi^{(1)} \otimes \Phi^{(1)}$ to rule out the absence of correlations.

\bigskip

Let us then devise a {\sc SuperQGAN} protocol to learn the $p_k^{(n)}$ of a general $n$-uses Pauli channel.
The Generator (G) agent tunes a fake distribution $q_{\Vec{k}}^{(n)}$ to generate its copy $\Phi_q^{(n)}$ of the channel in Eq. \eqref{pauli_ch_n}. Particularly, G will simulate the Pauli channel $\Phi_q^{(n)}$ by acting separately on the probes register with all the Pauli words appearing in Eq. \eqref{pauli_ch_n} and then weighting the results with the probabilities $q_{\Vec{k}}^{(n)}$.
For the sake of numerical simulations, the fake distribution will be parameterized with unbounded real parameters $\beta_{k_i^{(n)}}$ as
\begin{equation}
    q_{k_i}^{(n)}(\Vec{\beta}) = \frac{e^{-\beta_{k_i^{(n)}}}}{Z}, \qquad \text{with}\quad Z=\sum_{\Vec{k}^{(n)}} e^{-\beta_{\Vec{k}^{(n)}}}.
		\label{qbeta}
\end{equation}
This obviously introduces a redundant degree of freedom, but also allows us to discard any constraint issue on their domain.
The other agent, the Discriminator (D), will control both the measurement operator $M$ and the initialization circuit $I$. The measurement operator is modelled as a parameterized quantum circuit (PQC) \cite{benedetti2019parameterized} with parameters $\Vec{\theta_M}$, followed by a single-qubit measurement on the ancillary qubit M, see Fig.~\ref{fig:sens_qgan}(a). The initialization circuit is also modelled as a PQC with parameters $\Vec{\theta_I}$, entangling the system with ancillary qubits A. 
The resulting score function reads
\begin{equation}\label{score_sqgan}
\begin{aligned}
S(\Vec{\theta_I}, \Vec{\theta_M}, \Vec{\beta}) &= \tr\left[ M(\Vec{\theta_M})\left(\Phi_p^{(n)}(\rho(\Vec{\theta_I}))-\Phi_{q(\Vec{\beta})}^{(n)}(\rho(\Vec{\theta_I}))\right)\right] \\
&\,\,\hat{=}\,\, S_p - S_q
\end{aligned}
\end{equation}
where $M(\Vec{\theta_M})$ is the POVM element and $\rho(\Vec{\theta_I})$ is the global input state of the channels, both controlled by D.
The linearity of the trace allows us to rewrite each of the two terms $S_p$ and $S_q$ in \eqref{score_sqgan} as a sum of terms weighted by the respective distributions.
Suppressing parameters dependencies and indexes we may write $S_p = \sum_{\Vec{k}} p_{\Vec{k}}^{(n)} S(\sigma_{\Vec{k}})$ and analogously for $S_q$, with $S(\sigma_{\Vec{k}})$ being the scores associated to the Pauli string $\sigma_{\Vec{k}}$. We refer the reader to the Methods section \ref{sec:exp_setup} for a detailed description of the circuits architectures.
\begin{figure}[t!]
    \centering
		\includegraphics[width=1.\linewidth]{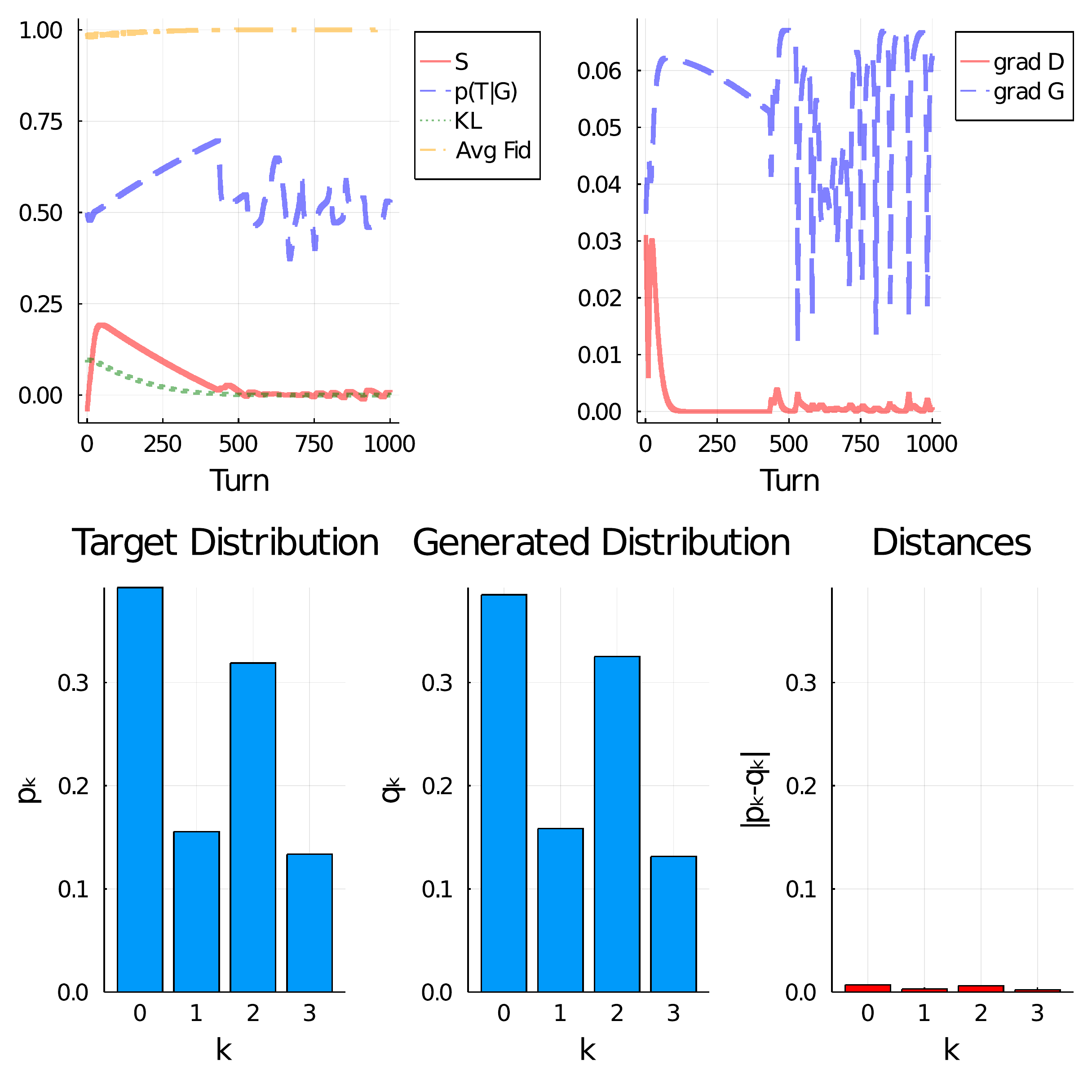}
		\caption{{\sc SuperQGAN} learning a single-use Pauli channel. Top panel shows training figures of merit (left), and gradients (right). Bottom one compares target and learnt distributions.\\
			The figures of merit that were tracked during training are: $S$, the score function \eqref{score_sqgan}; $p(T|G)$, G's objective function; KL, the Kullback-Leibler divergence between target and generated distributions; Avg Fid, the averaged fidelity between target and generated channels. The latter two quantities are defined in the main text above Eq. \eqref{MacPalmCorr}.
		}
    \label{fig:pc_spatial_1use}
\end{figure}
\begin{figure}[t!]
    \centering
		\includegraphics[width=1.\linewidth]{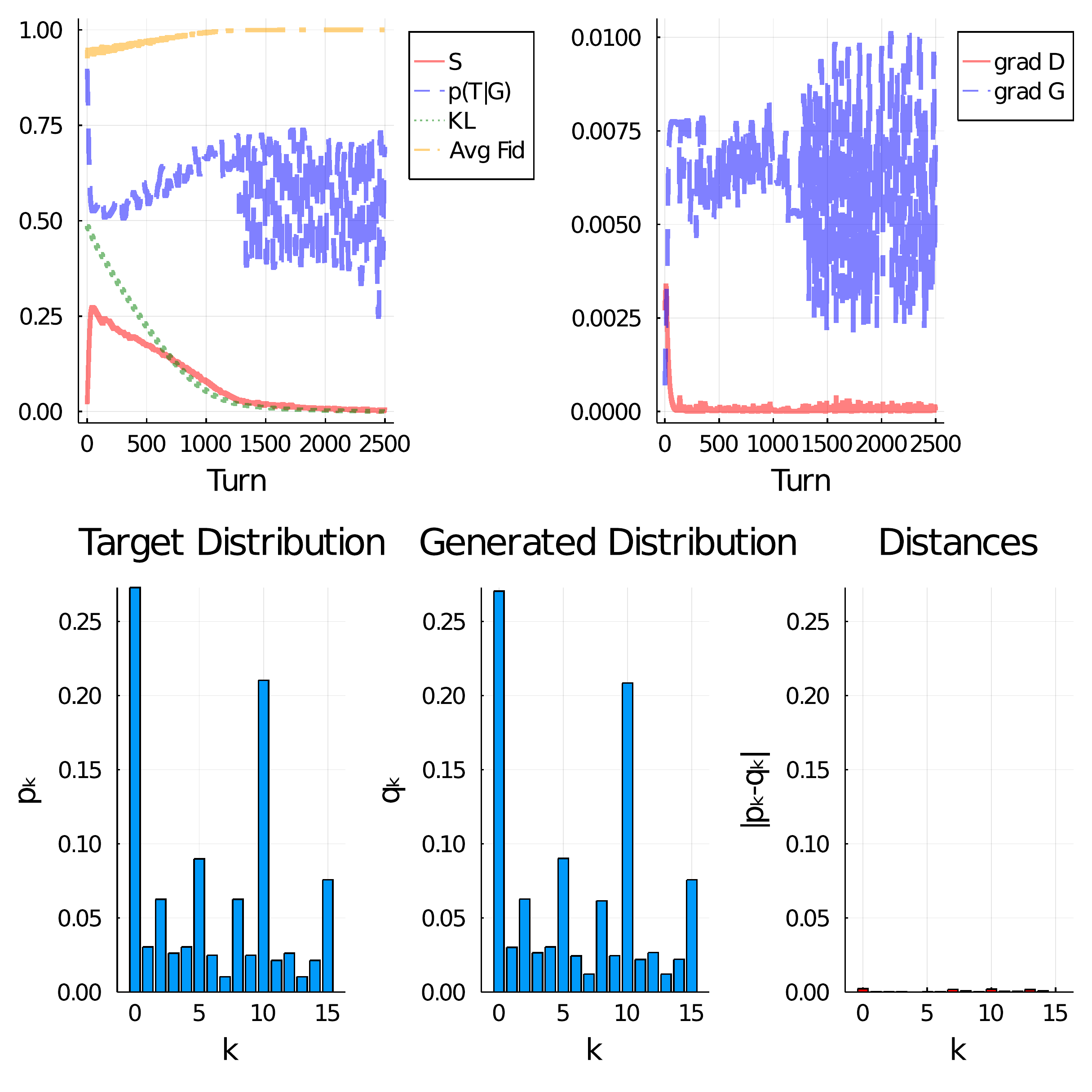}
		\caption{{\sc SuperQGAN} learning a two-uses spatially correlated Pauli channel. Top panel shows training figures of merit (left), and gradients (right). Bottom one compares target and learnt distributions, as described in Fig.~\ref{fig:pc_spatial_1use}. 
		The target distribution is generated using the single use distribution of Fig.~\ref{fig:pc_spatial_1use}, with the correlation law \eqref{MacPalmCorr} with $\mu=0.5$.
		}
    \label{fig:pc_spatial_2uses}
\end{figure}
\begin{figure}[t!]
    \centering
		\includegraphics[width=1.\linewidth]{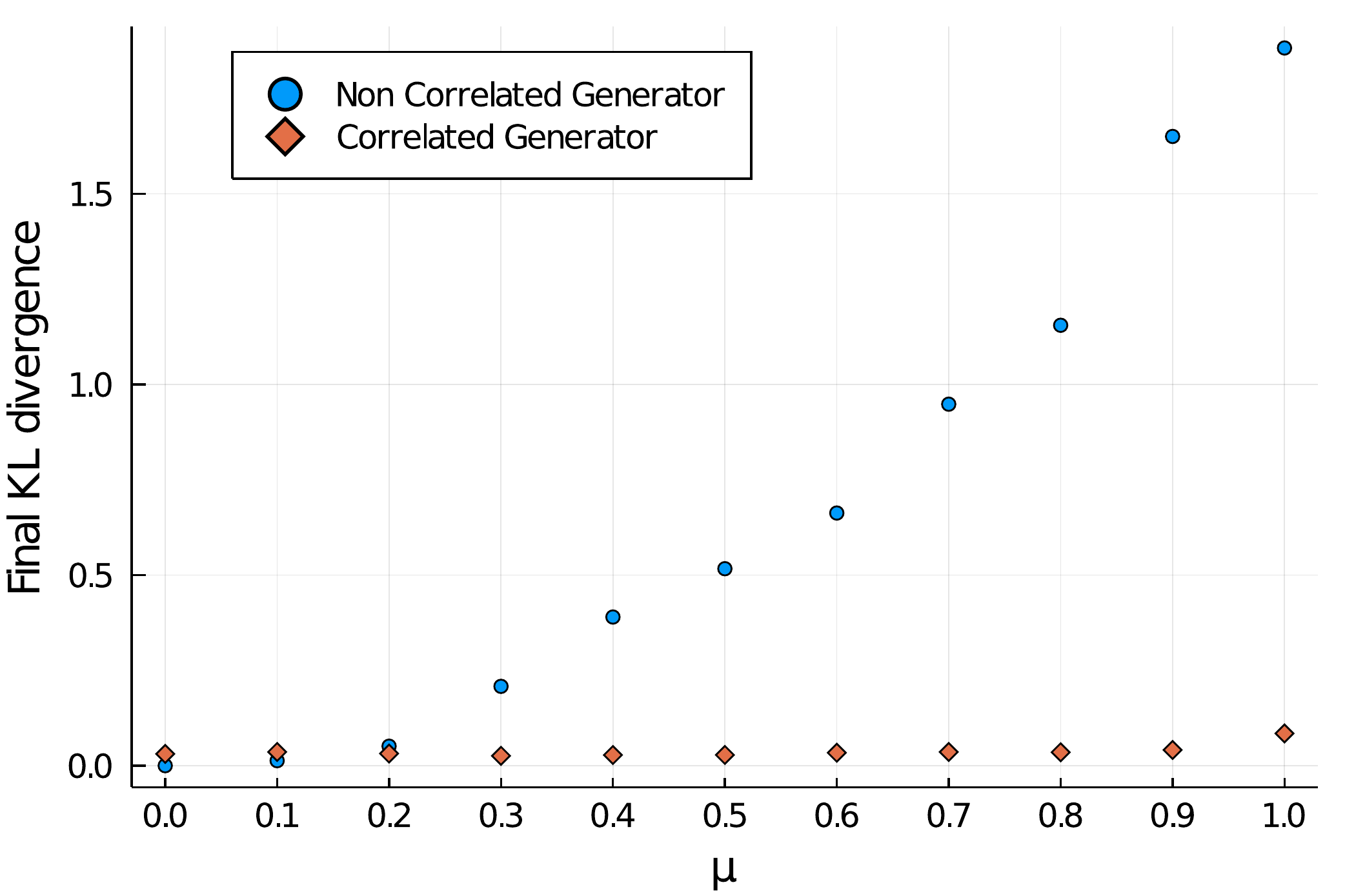}
		\caption{{\sc SuperQGAN} learning a correlated two-uses Pauli channel when G is not allowed to generate correlated distributions (blue dots) as opposed to the case when it can (red diamonds). Each point corresponds to a full learning procedure, where the target distribution is obtained from the prior of Fig.~\ref{fig:pc_spatial_1use} with the correlation of Eq. \eqref{MacPalmCorr}.
		}
    \label{fig:pc_spatial_nonCorrGenFailure}
\end{figure}
All our simulations are based on the Yao.jl quantum computation package for Julia \cite{luo2020yao}. 

In Fig. \ref{fig:pc_spatial_1use} we show the success of our protocol in assessing a single-use ($n=1$) Pauli channel. Among different figures of merit used to track the learning process, we stress the role of the Kullback-Leibler divergence and the averaged fidelity \cite{nielsen2010quantum}. The first one is a measure of similarity between two probability distributions $KL(p,q)=\sum_k p_k\log(p_k/q_k)$, whereas the second one is a distance measure over the space of quantum maps, given by the mean of the quantum fidelity $F(\rho, \sigma)=\left(\tr[\sqrt{\sqrt{\rho}\sigma\sqrt{\rho}}]\right)^2$ over the output states of the given maps when the input states are a basis of the corresponding Hilbert space.

In order to now test the performance of our setup when correlations may be present, we consider the multi-use scenario and we resort to a particular form of spatial correlations described by 
\begin{equation}\label{MacPalmCorr}
    p_{ij} = (1-\mu)p_i p_j +\mu p_i \delta_{ij}
\end{equation}
which has been introduced in \cite{macchiavello2002entanglement}, and interpolates between non-correlated channel for $\mu=0$ and a maximally correlated one for $\mu=1$.\\
When the generator is allowed to tune a generic distribution, the {\sc SuperQGAN} is able to learn the target distribution no matter the amount of correlations. In Fig. \ref{fig:pc_spatial_2uses} we show it with a two-uses example.

On the other hand, if G is constrained to generate non-correlated distributions only, i.e. is it is allowed to only tune a prior $q_k^{(1)}$ and use it to output $q_{\Vec{k}}^{(n)}=\prod_{j=1}^n q_{k_j}^{(1)}$, the process fails. Particularly, as one would expect, in this case the final Kullback-Leibler divergence grows with $\mu$, as shown in Fig.~\ref{fig:pc_spatial_nonCorrGenFailure}.

\subsection{Pauli channels: temporal correlations}\label{sec:temporal}

The interaction between the system and the surrounding environment typically gives rise to a
non-Markovian evolution of the system \cite{rivas2012open}. If the system is probed at 
discrete times $t_n$, such evolution can be expressed as a quantum comb
\cite{pollock2018non}, as in Fig.~\ref{fig:sens_qgan}b. 

Here we consider a simplified non-Markovian noise model based on Pauli channels with 
probability vectors $p^{(n)}_{\vec k}$, as in Eq.~\eqref{pauli_ch_n}.  While
in Eq.~\eqref{pauli_ch_n} the probabilities $p^{(n)}_{\vec k}$ describe the (possibly spatially 
correlated) noisy operations on different qubits, here $p^{(n)}_{\vec k}$  model the noisy 
operations on a {\it single} qubit but at different times, i.e.
\begin{equation}\label{pauli_ch_t}
	\Phi^{n}(\rho) = \sum_{\Vec{k}} p_{\Vec{k}}^{(n)} \sigma_{\Vec{k}} \rho \sigma_{\Vec{k}},
\end{equation}
where now $\sigma_{\Vec{k}}=\sigma_{k_1}\sigma_{k_2}\cdots\sigma_{k_n}$. 
In other terms, in Eq.~\eqref{pauli_ch_n} $k_s$ refers to the Pauli operation applied to the $s$-th qubit,
while in Eq.~\eqref{pauli_ch_t} $k_t$ refers to the Pauli operation applied to a single qubit at the $t$-th 
discrete iteration. One way to express the above circuit as the comb of Fig.~\ref{fig:sens_qgan}b is to 
assume that during the $t$-th iteration, and for all iterations $t=1,\dots,n$, 
the environment is measured with a four-outcome POVM, and depending 
on the measurement outcome $k_t$ a Pauli operation $\sigma_{k_t}$ is applied onto the system. 
The probability vector then models the joint probability of all possible POVM outcomes. 
If noise is temporally uncorrelated, then the probability is factorized 
$p_{\vec k}^{(n)}=\prod_{j=1}^n p({k_j})$. If noise is Markovian, then 
$p_{\vec k}^{(n)}=p(k_1)\prod_{j=2}^{n} p(k_j|k_{j-1})$. For more general probabilities, 
this model allows to describe non-Markovian noise. 

In a similar way to what we have done for spatial correlations, 
we can define a {\sc SuperQGAN} to learn $p_{\Vec{k}}^{(n)}$: as usual,
G tries to reproduce the distribution and D tries to discriminate between the real noise and 
the generated one. 
Although spatial and temporal Pauli correlations can both be modelled via $p_{\Vec{k}}^{(n)}$, the 
discrimination strategy can be entirely different. Indeed, the most general discrimination strategy 
for temporal correlations is the one depicted in Fig.~\ref{fig:sens_qgan}, where D inserts some 
probing operations at the intermediate times $t=1,\dots,n$ and, depending on the outcomes, 
decides whether the noisy channel was real or generated. In general, the probe alters the state, 
due to the wavefunction collapse, thus influencing all the future evolution.

\begin{figure}[h!]
    \centering
		\includegraphics[width=1.\linewidth]{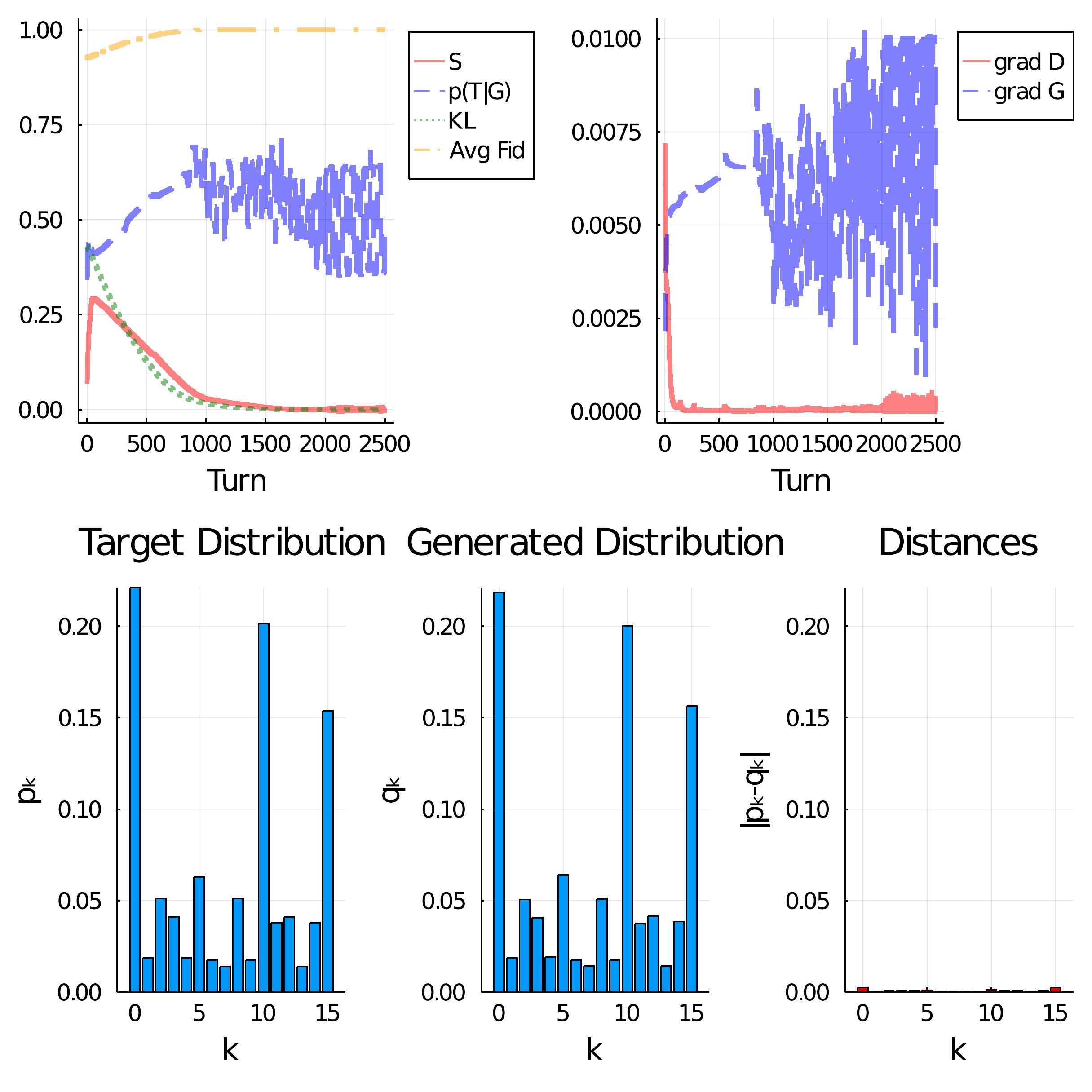}
		\caption{{\sc SuperQGAN} learning a two-uses temporally correlated Pauli channel. Top panel shows training figures of merit (left), and gradients (right). Bottom one compares target and learnt distributions, as described in Fig.~\ref{fig:pc_spatial_1use}.
		The target distribution is generated using a random single-use prior, using the correlation law \eqref{MacPalmCorr} with $\mu=0.5$.
		}
    \label{fig:pc_comb_2uses}
\end{figure}

\begin{figure}[h!]
    \centering
		\includegraphics[width=1.\linewidth]{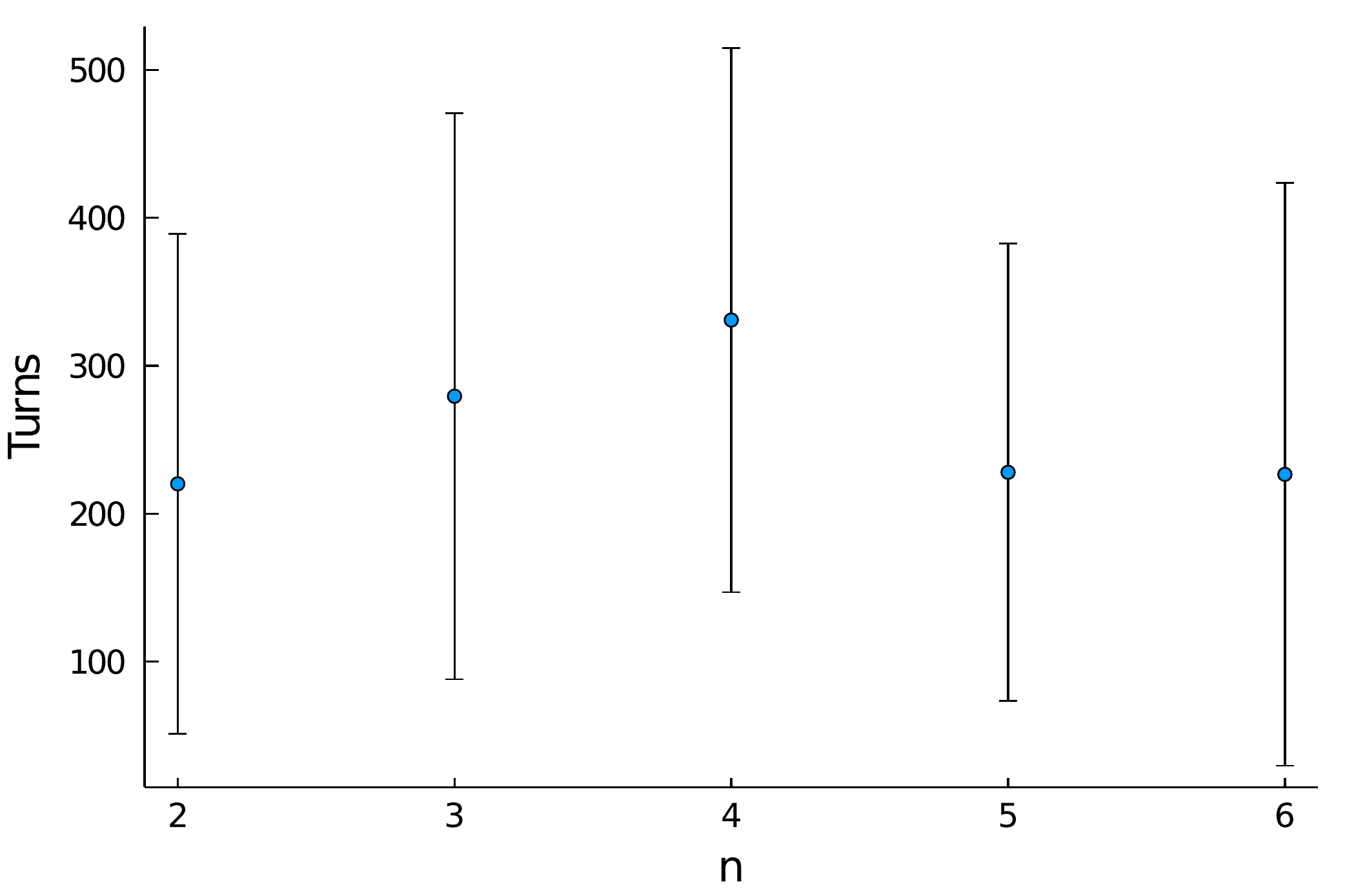}
		\caption{Total number of turns needed to achieve averaged fidelity greater than threshold value of 0.999 between target and generated channels. Each dot correspond to the mean over 10 runs of the modified {\sc SuperQGAN} whose generator knows the correlation model of Eq. \eqref{MacPalmCorr} and the $n=1$ probabilities ${p}$. Although the sample size is small, we observe that the number of iterations to achieve convergence does not increase with the number $n$ of channel uses, hence supporting the successful feasibility of our protocol for larger $n$.
		}
    \label{fig:iters_vs_n}
\end{figure}

We thus proceed to test the {\sc SuperQGAN} in the temporal correlations case. Again, the interested reader may find a detailed description of the {\sc SuperQGAN} architecture, as well as of the training procedure, in the Methods section.
In the single-use case the setup coincides with the one exploited for spatial correlations, as do the results. Hence, we analyze a correlated example given again in terms of the correlation law \eqref{MacPalmCorr}, which can be readily interpreted as a time correlation once the multi-indices are treated as time labels.
Particularly, in Fig.~\ref{fig:pc_comb_2uses} we show the success of our protocol in a two-uses Pauli channel.
Assuming to know in advance the model of correlation occuring in the quantum map, we can devise G in such a way that it only has to tune the correlations parameters, rather than the whole distribution. Then, we have used the learnt $n=1$ error rates ${p}$ to tackle the $n=(2,3,4,5,6)$ temporally correlated Pauli channels with a generator that only controls $\mu$. The number of turns needed to achieve convergence is found to be independent from $n$, as shown in Fig. \ref{fig:iters_vs_n}.

\subsection{Quantum metrology}\label{sec:noise_metr}

Quantum metrology \cite{giovannetti2011advances} can be rephrased as a {\sc SuperQGAN} with
$\delta$-like probability distribution in Eq.~\eqref{random_unitary_maps}, i.e.
$p(s) = \delta(s-\bar s)$. In other terms, we have a mapping implementing a unitary evolution 
$\rho\to U(\bar s)\rho U(\bar s)^\dagger$ and the metrology task is to estimate $\bar s$. 
Efficient quantum algorithms that fully exploit quantum effects to maximize the estimation precision 
typically employ either adaptive strategies 
or parallel applications of the unitary channel $U(\bar s)^{\otimes n}$ on an entangled state. 
Similar strategies are also needed when the parameter $s$ to be estimated 
is not fixed, but rather distributed according to some probability $p(s)$. 

In particular, here we consider a paradigmatic model of quantum metrology, i.e. the Mach-Zehnder-type interferometer, whose unitary evolution can be written as 
\begin{equation}
	U(s) = \begin{pmatrix}
		1 & 0 \\ 0 & e^{2\pi i s}
	\end{pmatrix}.
	\label{us}
\end{equation}
We assume that we can exactly express the parameter $s$ by  using $m$-bits as $s=\sum_{j=1}^m s_j/2^j$, 
where $0\leq s<1$ and $s_j\in\{0,1\}$, i.e. $s\equiv s_b = b/2^m$ for an integer $b<2^m$. 
When this assumption is not satisfied, we may get a reconstruction error. For instance, let us suppose to run the phase 
estimation algorithm for general $s$ using an $m+1$ qubit register. If $s_b$ is the best $m$-bit approximation of 
$s$, then the  algorithm will output $b'\neq b$ 
with probability $p_r(b'|b) = |2^{-m} (1-e^{2^m i \delta})/(1-e^{i\delta})|^2$, where $\delta=2\pi(s-s_b-s_{b'})$
\cite{nielsen2010quantum}. The distribution $p_r(b'|b)$ is peaked around $b'=b$ or around $b'=b\pm1$ when $2^ms$ is close 
to two different integers, so the reconstruction error is small and mostly limited to nearby values. In our analysis,
we fix $m$ and consider the error due to the finite $m$ as an imperfect reconstruction of $p(s)$. 

The number $n$ of independent applications of $U(s)$
needed to reconstruct $s$ with $m$-bit precision increases with $m$ \cite{hassani2017digital}. 
To simplify our treatment, here we assume that $m$ is {\it fixed}, so $p(s)$ becomes a discrete distribution with 
$2^m$ entries, and we consider $n$ parallel applications of $U(s)$. As a
result, we get the following random unitary channel
\begin{equation}
	\Phi_{R}^{(n)}(\rho) = \sum_{b=0}^{2^m-1} p(s_b) U(s_b)^{\otimes n} \rho\, U(s_b)^{\otimes n \dagger},
	\label{mz}
\end{equation}
where $s_b = b/2^m$ as above and $b$ is an integer.
The CJ state of each unitary channel $U(s)$ is a tensor product of a
maximally entangled pure state $\ket{\chi_s}^{\otimes n}$,
with $\ket{\chi_s} = (\ket{00}+2^{2\pi i s}\ket{11})/\sqrt 2$. To check for their linear independence, we may focus on the 
Gram matrix with the Hilbert-Schmidt product, $G_{st} = \tr[\chi_s^{\otimes n}\chi_t^{\otimes n}] = |\tilde G_{st}|^2$, 
where $\chi_s = \ket{\chi_s}\!\bra{\chi_s}$ and $\tilde G_{st} = 
\bra{\chi_s}\chi_t\rangle^{n}$. The Gram matrix has zero determinant, and hence at least a 
zero eigenvalue, when the matrices  $\chi_s^{\otimes n}$ are linearly dependent. 
The matrix $\tilde G$ can be diagonalized via a discrete Fourier transform, 
obtaining the eigenvalues $\tilde g_k = 2^{m-n} \sum_{\ell=0}^n \binom
n\ell\delta^{(2^m)}_{\ell,k}$ where $\delta_{ab}^{(c)}$ is 1 
if $a=b~({\rm mod})~c$ and 0 otherwise, and $k=0,\dots,2^m-1$. 
The eigenvalues of $G$ are then obtained via convolution 
$g_k = 2^{-m} \sum_u \tilde g_u \tilde g_{k\oplus u} = 2^{m-2n}\sum_\ell\binom n\ell \binom n{\ell\oplus k} $,
where $\oplus$ is the addition modulo $2^m$ and $\binom nk =0$ for $k>n$. 
Therefore, when $n<2^m/2$ at least one eigenvalue $g_k$ is zero and, accordingly, the mapping 
\eqref{mz} is not injective, namely two channels 	$\Phi_{R}^{(n)}$ may be equal even with different 
distributions $p(s)$. According to our analysis, 
 we need a number of probes satisfying
\begin{equation}
	n\geq 2^{m-1},
	\label{mzcond}
\end{equation}
to be sure that the reconstruction of $\Phi_{R}^{(n)}$ allows a unique reconstruction of $p(s)$. 

We now perform a numerical study using a {\sc SuperQGAN}, where G tries to generate a fake channel
with the same mathematical form of Eq.~\eqref{mz}, but different probability $q(s)$ instead of $p(s)$. 
To simplify the numerical treatment, G parametrizes its distribution again as in Eq.~\eqref{qbeta},
\begin{align}
	q(s) &= e^{-\beta_s}/Z, & Z=\sum_s e^{-\beta_s},
\end{align}
with real parameters $\beta_s$. 
The {\sc SuperQGAN} setup is analogue to that of spatial correlation learning outlined in Sec. \ref{sec:spatial} although we do not need to test all possible combination of unitaries since only tensor products appear. No differences in training performance are expected, and indeed, when one has enough resources, namely when \eqref{mzcond} is satisfied, G is always able to learn the correct distribution $p(s)$. In table \ref{tab:mvsn_mz} we show the final Kullback-Leibler divergence between $p(s)$ and $q(s)$ after the averaged fidelity between real and generated channels has reached the threshold value $f_{\text{tr}}=0.99999$. As one can see, sub-optimal values of $n$ lead to a learnt distribution $q(s)$ that is not converging to the target one.
\begin{table}[h!]
    \centering
    \begin{tabular}{||c || c | c ||} 
    \hline
    \multicolumn{3}{|c|}{} \\
    \hline
    \diagbox{$m$}{$n$} & $2^{m-1}$ & $2^{m-1} -1$ \\ [0.5ex] 
    \hline\hline
    2 & 0.000065(4) & 0.088(2) \\ 
    \hline
    3 & 0.00016(1) & 0.038(1) \\
    \hline
    4 & 0.00012(8) & 0.0015(1) \\ [1ex] 
    \hline
    \end{tabular}
    \caption{Final values of Kullback-Leibler divergence $KL(p,q)$ between target distribution $p(s)$ appearing in Eq. \eqref{mz} and G's generated one $q(s)$. The {\sc SuperQGAN} is stopped as soon as the averaged fidelity between target and fake channels gets larger than $0.99999$. Values reported here refer to an average over $10$ runs with fixed targets and different (random) parameters' initializations.}
    \label{tab:mvsn_mz}
\end{table}

\section{Conclusions}\label{sec:conclusions}

In this work we generalize standard Quantum Generative Adversarial Networks for quantum state learning to what we here call {\sc SuperQGAN}s, i.e.~a quantum machine learning tool to learn and reproduce quantum maps.
Particularly, we address the case of Pauli channels with a method that easily extends to more general random unitary maps. Exploiting QGANs ability of dealing with mixed quantum states when a particular class of optimizing algorithms are employed, we are able to show how the {\sc SuperQGAN} setup is capable of learning Pauli channels with different types and amount of correlations.

Even though the actual setup of the {\sc SuperQGAN} changes when spatial or temporal correlations are involved, our method always relies on parameterized quantum circuits to model generative and discriminative agents that compete against each other until the former learns to exactly reproduce the quantum map at hand and fools the latter. The ability to separately tackle temporal or spatial correlations allows us to better classify the unwanted couplings that unavoidably spoil nowadays quantum computations in the NISQ devices. As shown in Ref. \cite{wallman2016noise}, noise affecting quantum processors can be controlled in such a way to be effectively described by Pauli channels via the implementation of randomized compiling. Thus, having automatic methods such as {\sc SuperQGAN}s to characterized the latter will help devise optimal error mitigation protocols.
Without any constraints, the {\sc SuperQGAN} implements full quantum process tomography, which obviously scales exponentially with $n$, be it the number of probes in a spatially correlated configuration or the number of successive uses for a temporal one. However,  more interestingly, the analysis that led to Fig. \ref{fig:iters_vs_n} shows that when the noise model is constrained to a given form, such as when we already have some insight on the type of correlations affecting the device at hand, the {\sc SuperQGAN} method is efficient and the resources it needs do not scale with $n$.

In addition, following the analysis of the teleportation-induced correlated quantum channels in Ref. \cite{caruso2010teleportation}, once our new protocol learns the probabilities $p^{(n)}_{\vec k}$ of the Pauli channels, one can also analytically calculate the corresponding quantum capacities (known only in a very few cases) and the distillable entanglement of a generic bipartite quantum state being exploited to implement a teleportation protocol that can be always mapped to a correlated Pauli channel. Indeed all these quantities are very simple analytical functions of the probabilities $p^{(n)}_{\vec k}$
\cite{caruso2010teleportation,pirandola2017fundamental}. Therefore, these results are expected to find applications also in other fields other than quantum computing, as quantum communication and quantum cryptography. Notice, for instance, that quantum error correction and quantum teleportation are indeed the crucial building blocks towards the feasible realization of the so-called Quantum Internet \cite{wehner2018quantum}.

Finally, we also show how the {\sc SuperQGAN} machinery fits the problem of quantum metrology, i.e. the estimation of a parameter upon which a certain quantum map depends. We apply it to the estimation of the phase shift $s$ induced by a Mach-Zehnder interferometer, and find the optimal setup under which $s$ can be faithfully reconstructed. 
We believe that quantum metrology via {\sc SuperQGAN}s will be particularly useful for those scenarios when the theoretically optimal entangled input state is not known, since our method has the ability to cleverly combine input preparation and final measurement to find the best discrimination policy.

\begin{acknowledgements}
	L.B. acknowledges support by the program ``Rita Levi Montalcini'' 
	for young researchers, Grant No. PGR15V3JYH,  funded by 
	``Ministero dell'Istruzione, dell'Universit\`a e della Ricerca (MIUR)''. F.C. was financially supported by the Fondazione CR Firenze through the project QUANTUM-AI, the PATHOS EU H2020 FET-OPEN Grant No. 828946, and the Florence University Grant Q-CODYCES.
\end{acknowledgements}

\section*{Methods}

\subsection*{SuperQGAN setup}\label{sec:exp_setup}

\begin{figure}[h!]
    \centering
		\includegraphics[width=1.\linewidth]{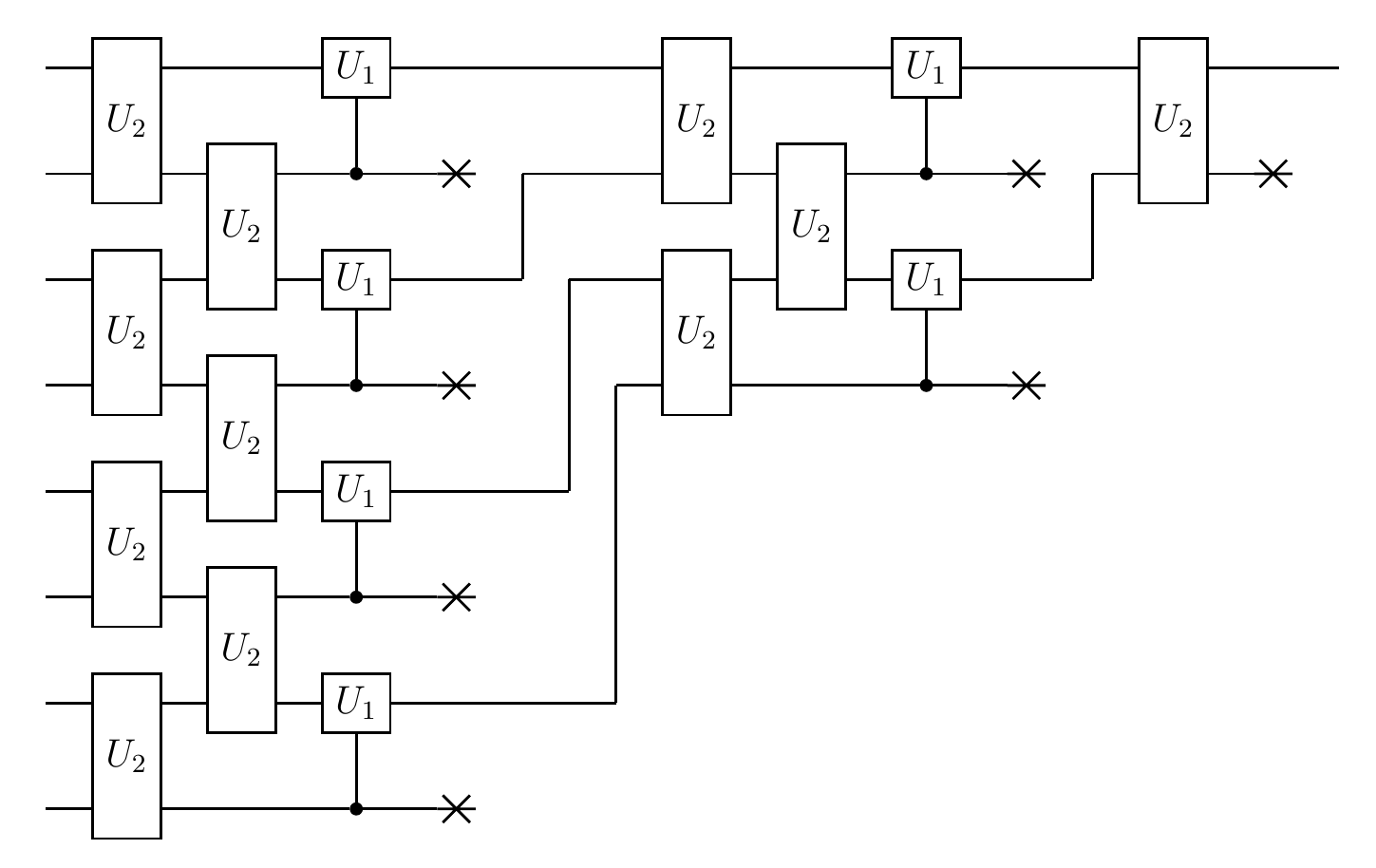}
		\caption{An example of QCNN. $U_1$ and $U_2$ are elements of $SU(2)$ and $SU(4)$, respectively, and their circuital realization is explained in Fig. 3 of \cite{braccia2020enhance}. Crossed lines stand for \textit{forgotten} qubits. Notice that the latter do not get measured, and are to be kept untouched until the end of the network. 
		}
    \label{fig:qcnn_struct}
\end{figure}

Here we describe the technicalities of the numerical simulations described in the main text.
First of all let us introduce the building blocks of the parametric quantum circuits used in both the spatial and temporal configurations.
These are basically two: a two-qubit generic $SU(4)$ operation, obtained with the recipe described in \cite{shende2004minimal}, and the \textit{quantum convolutional neural network} (QCNN) introduced in \cite{cong2019quantum} and illustrated in Fig. \ref{fig:qcnn_struct}. Particularly, the final PQC controlled by D to implement its POVM is a QCNN, whereas intermediate operations are composed by alternating layers of $SU(4)$.

Another aspect that is common to both configurations is gradients evaluation. Indeed, we have always resorted to the \textit{parameter shift rule} described in \cite{schuld2019evaluating,banchi2021measuring}.\\
Lastly, as far as the optimization procedure is concerned, we have used a variation of the \textit{Optimistic Mirror Descent} (OMD) introduced in \cite{daskalakis2017training} and fruitfully tested in \cite{braccia2020enhance}, namely \textit{Optimistic ADAM}. This optimization strategy uses ADAM \cite{kingma2014adam}, i.e. a famous gradient descent with momentum optimizer, to evaluate the parameters increment at step $t$, i.e. $\delta_t$, then implements it as in OMD $\theta \leftarrow \theta -2\eta\delta_t + \eta\delta_{t-1}$, where $\eta$ is the learning rate and $\theta$ is the parameter being updated. \\
Hyperparameters such as the agents' learning rates, the total number of training turns and the single agents' number of updating steps are tuned case by case. There is, however, a rule of thumb: learning rates $\eta$ are chosen inside the range $0.01<\eta<0.25$ and D's (and I's) steps are almost always twenty times more than G's ones.

\subsection{Spatial correlations}

The circuits architecture of the {\sc SuperQGAN} for spatially correlated channels is shown in Fig. \ref{fig:sens_qgan}(a).
As discussed before, I is built of staggered layers of $SU(4)$ blocks and its depth is hand-tuned depending on the number of qubits it acts on. G simulates the random unitary channel at hand by separately applying the unitaries defining it and later weighting the discriminator outcomes with its parametric distribution $q_k$. The discriminator evaluates $p(R|\cdot)$ applying its QCNN on the output state of the channel being tested, R or G, and on its ancilla qubit. The role of the QCNN is to filter and encode the relevant information in the ancilla qubit, so that measuring it D can tell the difference between real and generated data.

\subsection{Temporal correlations}

When tackling temporal correlations, the {\sc SuperQGAN} adopts a comb-like architecture, as shown in Fig. \ref{fig:sens_qgan}(b).
Now I and D have access to an extra \textit{workspace} qubit, which they will use to store information and process the temporal memory of the channel being tested. Now, after the initial state preparation implemented by I, whose structure is the same as the previous setup, D acts after each channel use with a similar PQC, and only after the last use implements the measurement via a QCNN.

\bibliography{sensingbib}

\end{document}